\documentclass{emulateapj}

\shorttitle{Nucleosynthesis in Electron Capture Supernovae}

\shortauthors{Wanajo et al.}

\begin{document}


\title{Nucleosynthesis in Electron Capture Supernovae of AGB Stars}

\author{S. Wanajo\altaffilmark{1, 2},
        K. Nomoto\altaffilmark{1, 2},
        H. -T. Janka\altaffilmark{3},
        F. S. Kitaura\altaffilmark{3},
        and B. M\"uller\altaffilmark{3}}

\altaffiltext{1}{Institute for the Physics and Mathematics of the Universe, 
       University of Tokyo, Kashiwa, Chiba 277-8582, Japan}

\altaffiltext{2}{Department of Astronomy, School of Science,
        University of Tokyo, Bunkyo-ku, Tokyo, 113-8654, Japan;
        wanajo@astron.s.u-tokyo.ac.jp, nomoto@astron.s.u-tokyo.ac.jp}

\altaffiltext{3}{Max-Planck-Institut f\"ur Astrophysik,
        Karl-Schwarzschild-Str. 1, D-85748 Garching, Germany;
        thj@mpa-garching.mpg.de, kitaura@mpa-garching.mpg.de}

\begin{abstract}
We examine nucleosynthesis in the electron capture supernovae of
progenitor AGB stars with an O-Ne-Mg core (with the initial stellar
mass of $8.8\, M_\odot$). Thermodynamic trajectories for the first
810~ms after core bounce are taken from a recent state-of-the-art
hydrodynamic simulation. The presented nucleosynthesis results are
characterized by a number of distinct features that are not shared
with those of other supernovae from the collapse of stars with iron
core (with initial stellar masses of more than $10\, M_\odot$). First
is the small amount of $^{56}$Ni ($= 0.002-0.004\, M_\odot$) in the
ejecta, which can be an explanation for observed properties of faint
supernovae such as SNe~2008S and 1997D. In addition, the large Ni/Fe
ratio is in reasonable agreement with the spectroscopic result of the
Crab nebula (the relic of SN~1054). Second is the large production of
$^{64}$Zn, $^{70}$Ge, light $p$-nuclei ($^{74}$Se, $^{78}$Kr,
$^{84}$Sr, and $^{92}$Mo), and in particular, $^{90}$Zr, which
originates from the low $Y_e$ ($= 0.46-0.49$, the number of electrons
per nucleon) ejecta. We find, however, that only a $1-2\%$ increase of
the minimum $Y_e$ moderates the overproduction of $^{90}$Zr. In
contrast, the production of $^{64}$Zn is fairly robust against a small
variation of $Y_e$. This provides the upper limit of the occurrence of
this type of events to be about $30\%$ of all core-collapse
supernovae. 
\end{abstract}

\keywords{
nuclear reactions, nucleosynthesis, abundances
--- stars: abundances
--- supernovae: general
--- supernovae: individual (SN~1054, SN~1997D, SN~2008S)
--- nebulae: Crab Nebula
}

\section{Introduction}

Massive stars end their lives with core-collapse supernovae
(SNe~II/Ibc), which are the predominant sources of having enriched
galaxies with the elements heavier than helium. The other type,
thermonuclear supernovae (SNe~Ia), also contributes to the enrichment
of iron-peak elements, which is, however, absent in the early
universe. The metals produced by core-collapse supernovae serve as
diagnostic tools to uncover the chemical-enrichment history of the
Galaxy from its poorly understood early stage to the present day. A
reliable prediction of supernova yields has been, however, hampered by
the yet unknown mechanism that causes the explosion. Previous studies
of supernova nucleosynthesis have relied upon a number of model
parameters such as the explosion energy, the position that divides the
ejecta and the remnant (mass cut), and the electron fraction ($Y_e$,
the number of electrons per nucleon). The production of each element
in a supernova, in particular of those synthesized in the innermost
ejecta, is severely affected by the choice of these parameters
\citep[see, e.g.,][]{Tomi07, Hege08}. It is obvious that 
nucleosynthesis studies with self-consistently exploding models are
eventually needed to obtain reliable supernova yields.

Recent one-dimensional simulations including accurate neutrino
transport seem to exclude the possibility of neutrino-driven
explosions without the help of multi-dimensional effects
\citep{Ramp00, Mezz01, Lieb01, Thom03, Sumi05, Bura06}, at least for
standard nuclear and neutrino physics input. An exception are the
explosions of $8-10\, M_\odot$ stars. A star in this mass range forms
an electron-degenerate core consisting of oxygen, neon, and magnesium
(O-Ne-Mg) during the final stage of its evolution (instead of an iron
core in the case of more massive stars), and becomes a super
asymptotic giant branch (AGB) star. Such a star ends its life either
as an O-Ne-Mg white dwarf or a core-collapse supernova leaving behind
a neutron star. For the latter, the collapse is induced by electron
capture (what is called an ``electron capture supernova'') when the
core mass grows to $1.38\, M_\odot$ and the central density reaches
$4\times 10^9$~g~cm$^{-3}$. However, the uncertainties in mixing and
mass loss during the evolution make it difficult to draw a clear line
between these two channels \citep{Nomo84, Nomo87}. Recent studies
report that only the range close to the upper end \citep[a mass range
of $\lesssim 1\, M_\odot$ or $\sim 4\%$ of all
supernovae,][]{Sies07,Poel08} leads to the explosion channel, although
the range could be wider for lower metallicity stars.

The structure of the O-Ne-Mg core is distinctively different from the
iron cores of more massive stars by the fact that it has a steep
density gradient in the outermost layers, surrounded by an extremely
extended, loosely bound H/He envelope.  Recently, \citet{Kita06} have
obtained self-consistent explosions from the collapse of the O-Ne-Mg
core in a stellar progenitor with an initial mass of $8.8\, M_\odot$
developed by \citet{Nomo84}. Their one-dimensional simulations with a
state-of-the-art, energy-dependent treatment of the neutrino transport
are in fact the only recent models in the literature with successful
supernova explosions in spherical symmetry for standard nuclear and
weak interaction physics \citep[see also][for a similar
result]{Burr07}. The explosions are initiated by the rapid outward
acceleration of the supernova shock when it encounters the steep
density gradient and fast decline of the mass accretion rate at the
edge of the O-Ne-Mg core. They are powered by the neutrino-heating
mechanism, which yields a low explosion energy of $\sim$1--$2\times
10^{50}$~erg \citep[for a detailed discussion, see][]{Jank08a}. The
new calculations are a revision of previous hydrodynamic results for
the same O-Ne-Mg core, namely, of prompt explosions \citep{Hill84,
Wana03}, powerful neutrino-driven explosions \citep[$0.6-1.2\times
10^{51}$~erg,][]{Mayl88}, and no explosions \citep{Burr85, Baro87}.

The purpose of this paper is to present nucleosynthesis results for
the first 810~ms of the neutrino-driven explosion of a collapsing
O-Ne-Mg core (electron capture supernova), using the thermodynamic
trajectories obtained by \citet{Kita06}. Employing a self-consistently
calculated explosion model with sophisticated neutrino transport is of
particular importance not only for the nucleosynthesis study itself,
but also for a couple of other aspects.  On the one hand, the mass
range of 8--$10\, M_\odot$ accounts for about 30\% of all the
core-collapse supernova events, if all the range leads to the
explosion channel. The electron capture supernovae can thus be
potentially significant contributors to the Galactic chemical
evolution of some species. On the other hand, explosions from these
progenitors have been proposed as a possible explanation for the
inferred low explosion energy of SN~1054 \citep[Crab
supernova,][]{Nomo82, Davi82} \citep[see also][for the $10\, M_\odot$
progenitor with an iron core]{Hill82} as well as for the small
$^{56}$Ni amount estimated for some low-luminosity supernovae
\citep[e.g., SN~1997D,][]{Tura98, Chug00, Bene01, Hend05}. A newly
identified class of luminous transients like SN~2008S, whose
progenitors were deeply dust-enshrouded massive stars, is also
suggested to be electron capture supernovae \citep{Prie08,Thom08}
\citep[see also][]{Past07}.

Recently, \citet{Hoff08} have investigated the nucleosynthesis in
electron capture supernovae, using an explosion calculation of
\citet{Jank08a} that is very similar to the explosion models of
\citet{Kita06} \citep[but was computed with slightly different input
physics, see ][]{Jank08a}. Their study was aimed at determining
whether the conditions are suitable for an $r$-process. They found no
$r$-process formation and a severe overproduction of $^{90}$Zr, which
is also seen in our results. In our work, however, we perform
nucleosynthesis calculations in much more detail, taking into account
a number of possible uncertainties. In the following section (\S~2)
the explosion models of \citet{Kita06} and the methods for the
nucleosynthesis calculations will be described. The nucleosynthesis
results for the original model and for various modifications of it
will be presented in \S~3. We will discuss the question whether
electron capture supernovae can be significant contributors to
Galactic chemical evolution in \S~4 and will address the possibility
that they are the origin of some low-energy supernovae in \S~5.
Finally, our summary and conclusions will follow in \S~6.

\section{Explosion Model and Reaction Network}

\citet{Kita06} have simulated the collapse of the O-Ne-Mg core
\citep[with the mass of $1.38\, M_\odot$,][]{Nomo84} \citep[see
also][]{Miya80}, employing two different (``soft'' and ``stiff'')
nuclear equations of state (EoSs). In this study, we adopt the result
with the softer EoS \citep[][LS]{Latt91} as the ``standard'' model
(labelled ST), which is qualitatively similar to that with the stiffer
one \citep[WH; Wolff \& Hillebrandt EoS in][]{Hill84}. The model with
the WH EoS is used later for the purpose of comparison (\S~3.2).
Figure~1 shows the evolution of the radius, density, temperature, and
$Y_e$ for selected mass elements in model ST as a function of
post-bounce time $t_\mathrm{pb}$. The ejecta in model ST are split
into 29 mass shells with the first ejected zone having a mass of about
$4\times 10^{-4}\, M_\odot$ and each of the other ones of about
$5\times 10^{-4}\, M_\odot$ ($1.39\times 10^{-2}\,M_\odot$ in total).

\begin{figure}
\epsscale{1.0}
\plotone{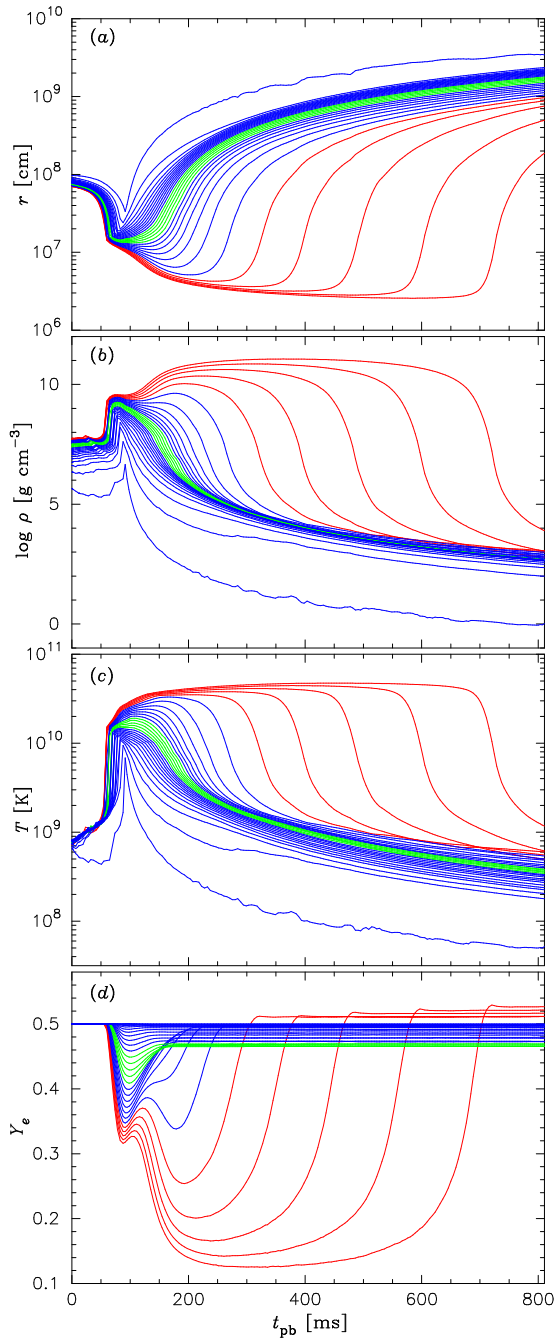}
\caption{Radius (\textit{a}), density (\textit{b}), temperature
(\textit{c}), and $Y_e$ (\textit{d}) as a function of post-bounce time
for material ejected from the collapsing O-Ne-Mg core in model ST. The
trajectories are colored in green, blue, and red for $Y_e < 0.470$,
$0.470 < Y_e < 0.500$, and $Y_e > 0.500$ (values at the end of
simulation), respectively.}
\end{figure}

The nucleosynthesis yields in each mass shell are obtained in a
post-processing step by solving an extensive nuclear reaction
network code. The network consists of 6300 species between the proton-
and neutron-drip lines predicted by the recent fully microscopic mass
formula \citep[HFB-9,][]{Gori05}, all the way from single neutrons and
protons up to the $Z = 110$ isotopes. All relevant reactions,
i.e. $(n, \gamma)$, $(p,\gamma)$, $(\alpha, \gamma)$, $(p, n)$,
$(\alpha, n)$, $(\alpha, p)$, and their inverses are included. The
experimental data, whenever available, and the theoretical predictions
for light nuclei ($Z < 10$) are taken from the
REACLIB\footnote{http://nucastro.org/reaclib.html.} compilation. All
the other reaction rates are taken from the Hauser-Feshbach rates of
BRUSLIB\footnote{http://www.astro.ulb.ac.be/Html/bruslib.html.}
\citep{Aika05} making use of experimental masses \citep{Audi03}
whenever available or the HFB-9 mass predictions \citep{Gori05}
otherwise. The photodisintegration rates are deduced from the reverse
rates applying the reciprocity theorem with the nuclear masses
considered. The weak and intermediate screening corrections to charged
particle reactions are adopted from \citet{Grab73}.

The $\beta$-decay rates are taken from the gross theory predictions
\citep[GT2,][]{Tach90} obtained with the HFB-9 predictions
(T. Tachibana 2005, private communication). Electron capture reactions
on free nucleons and on heavy nuclei \citep{Full82, Lang01} as well as
rates for neutrino capture on free nucleons and $^4$He and for
neutrino spallation of free nucleons from $^4$He \citep{Woos90,
McLa96} are also included. In contrast, neutrino-induced reactions of
heavy nuclei are not taken into account in this study, but they are
expected to make only minor effects \citep{Meye98b}. Figure~2 shows the
luminosities and mean energies for neutrinos of all types as functions
of the post-bounce time $t_\mathrm{pb}$ in model ST; these results are
taken from \citet{Kita06} and used for calculating the rates of
neutrino-induced reactions. More preciously, we should apply those
quantities in the co-moving frame of the fluid with corrections for
the gravitational redshift and for Doppler shifting due to fluid
motion.  We neglect such effects, which are important on the one hand
only when the fluid is relatively close to the neutrino sphere ($<
\textrm{several}\, 10$~km), where the temperature is still higher than
$T_9 = 9$ (see below). On the other hand, at large distances, where
the expansion velocities of the gas are larger, neutrino interactions
become essentially irrelevant.

\begin{figure}
\epsscale{1.0}
\plotone{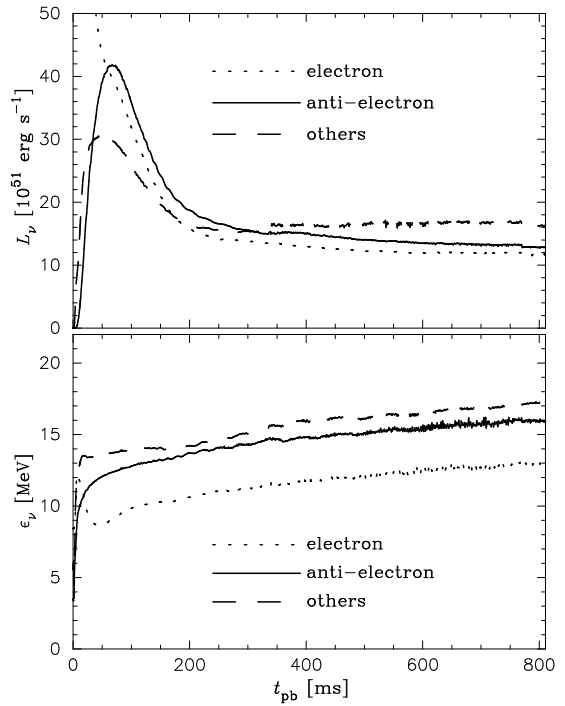}
\caption{Neutrino luminosities (\textit{top}) and mean neutrino energies
(\textit{bottom}) as functions of post-bounce time for electron
(\textit{dotted line}), anti-electron (\textit{solid line}), and 
heavy-lepton (\textit{dashed line}) neutrinos. The data are given
for an observer at rest at 400~km from the center.}
\end{figure}

Each nucleosynthesis calculation is initiated when the temperature
decreases to $T_9 = 9$ (where $T_9 \equiv T/10^9\, \mathrm{K}$). In
the first ejected trajectory, the highest temperature is $T_9 \approx
7$ (Fig.~1), which is taken to be the initial condition for this case
only. At such high temperatures, the composition is in the nuclear
statistical equilibrium (mostly free nucleons and few
$\alpha$ particles), which is realized 
immediately after the calculation starts. The
initial compositions is then given by $X_n = 1 - Y_{e,\mathrm{i}}$ and $X_p =
Y_{e,\mathrm{i}}$, respectively, where $X_n$ and $X_p$ are the mass fractions
of free neutrons and protons, and $Y_{e,\mathrm{i}}$ is the initial electron
fraction at $T_9 = 9$ (Fig.~3, \textit{black line} for model ST).

\begin{figure}
\epsscale{1.0}
\plotone{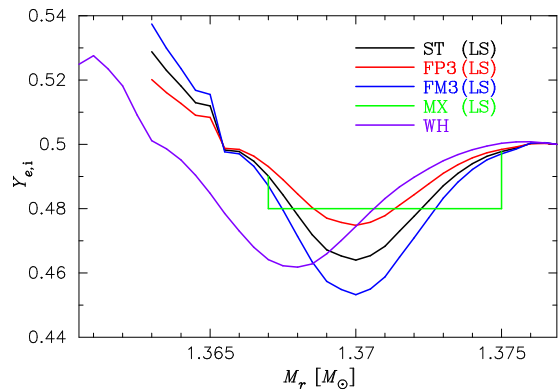}
\caption{Initial electron fraction $Y_{e,\mathrm{i}}$ for material
ejected from the core as a function of the enclosed mass,
$M_r$. Different colors correspond to models ST, FP3, FM3, MX, and WH
as denoted in the panel (see text).}
\end{figure}


\section{Nucleosynthesis Results}

In subsection 3.1, we will present the nucleosynthesis results for 
the unmodified model (ST) of \citet{Kita06}, and the corresponding
information for variations of model ST will then be discussed
in the following subsections. In each model, the
nucleosynthetic yields for all the trajectories are mass-integrated
over the ejecta-mass range.

\subsection{Unmodified Model}

The nucleosynthesis results of the unmodified model (ST) are shown in
Figures~4 (isotopes) and 5 (elements). Both plots present the
overproduction factors defined by the mass fractions in the ejecta
with respect to their solar values \citep{Lodd03}. The even-$Z$ and
odd-$Z$ species are denoted by circles and triangles, respectively. In
Figure~4, the isotopes for a given element are connected by lines, and
the abundances smaller than $X/X_\odot < 40$ are omitted. The dotted
horizontal lines indicate a ``normalization band'' between the largest
production factor ($^{90}$Zr and Zr in Figs.~4 and 5, respectively)
and a factor of ten smaller than that, along with the median value
(\textit{dashed line}). This band is taken to be representative of the
uncertainty in the nuclear data involved. In the following, we
consider that electron capture supernovae can be contributors to the
solar (or Galactic) inventories of the species located within the
normalization band.

\begin{figure}
\epsscale{1.0}
\plotone{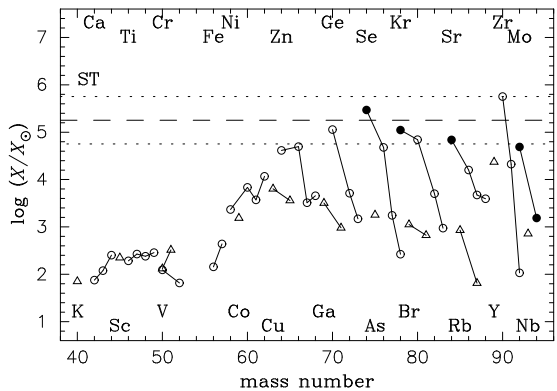}
\caption{Mass fractions of isotopes (after decay) in the ejecta of
model ST relative to their solar values \citep{Lodd03} as
functions of mass number. The abundances smaller than $X/X_\odot < 40$
are omitted. The even-$Z$ and odd-$Z$ isotopes are denoted by open
circles and triangles, respectively. The $p$-nuclei are represented
with filled symbols. The solid lines connect isotopes of a given
element. The dotted horizontal lines indicate a ``normalization band''
between the largest production factor and a factor of ten smaller than
that, along with the median value (\textit{dashed line}).}
\end{figure}

\begin{figure}
\epsscale{1.0}
\plotone{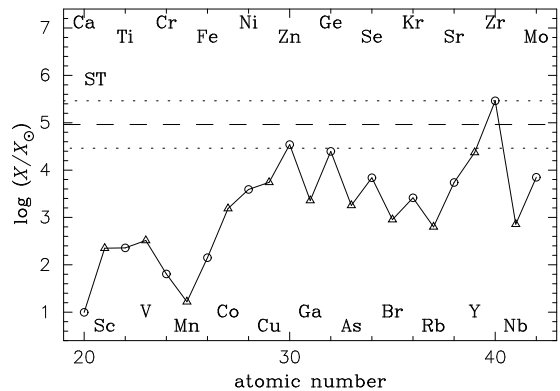}
\caption{Mass fractions of elements (after decay) in the ejecta of
model ST relative to their solar values \citep{Lodd03} as a
function of atomic number. The even-$Z$ and odd-$Z$ elements are
denoted by open circles and triangles, respectively.}
\end{figure}

Figure~4 indicates that, along with the marginal ones, model ST can
account for the production of $^{64, 66}$Zn, $^{70}$Ge, $^{74, 76}$Se,
$^{78, 80}$Kr, $^{84}$Sr, $^{90}$Zr, and $^{92}$Mo, where all the
light $p$-nuclei (\textit{filled circles}) up to $A=92$ are
included. However, only a few elements (Zn, Ge, Y, and Zr) fall into
the normalization band (Fig.~5), since the $p$-nuclei comprise only
small fractions of a given element (0.89\%, 0.35\%, 0.56\%, and 14.8\%
for $^{74}$Se, $^{78}$Kr, $^{84}$Sr, and $^{92}$Mo, respectively). Our
result for model ST is in reasonable agreement with that in
\citet[][Fig.~1]{Hoff08}\footnote{The total ejecta mass including the
outer H/He envelope is taken to be $1.263\, M_\odot$ in
\citet{Hoff08}, while we consider only the calculated zones with
$1.39\times 10^{-2}\, M_\odot$. This leads to the hundred times larger
values of $X/X_\odot$ in this paper.}, and some differences are due to
the slightly higher minimum $Y_e$ in our model (\S~3.5).

Figure~6 shows the mass fractions of some important isotopes (after
decay) for model ST (\textit{top}) and these mass fractions relative
to their solar values (\textit{bottom}) as functions of the enclosed
mass $M_r$. In each mass shell, the dominant heavy isotope is either
$^{56}$Fe (produced in the form of $^{56}$Ni) or $^{58,60}$Ni,
depending on $Y_{e,\mathrm{i}}$. 
In the neutron-rich ejecta, the nuclear products in the single
quasi-statistical-equilibrium (QSE) cluster have peaks at $A\sim 60$
and $A\sim 90$ because of the strong binding at $N = 28$ and 50
\citep{Meye98a}.  The abundance of $^{90}$Zr is maximal in the mass
shell with the lowest $Y_{e,\mathrm{i}} = 0.464$. In addition, the
matter with $Y_{e,\mathrm{i}} = 0.46-0.49$ is so \textit{proton-rich}
(compared to the average of the stable isotopes in the vicinity, e.g.,
the proton fraction of $^{90}$Zr is 0.44) that the successive proton
captures lead to the production of light $p$-nuclei up to $A =
92$. This can also be seen in the early neutrino-driven winds
\citep{Hoff96, Wana06}.

\begin{figure}
\epsscale{1.0}
\plotone{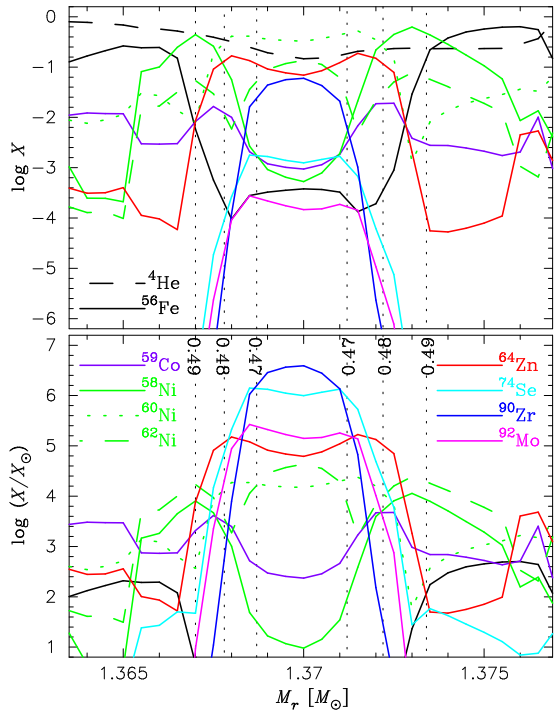}
\caption{Mass fractions of several important isotopes (after decay) in
the ejecta of model ST (\textit{top}), and these mass fractions relative
to their solar values (\textit{bottom}) as functions of the enclosed mass
$M_r$. The vertical dotted lines indicate the mass coordinates where the
initial $Y_{e,\mathrm{i}}$ is 0.470, 0.480, and 0.490 (see Fig.~3).}
\end{figure}

It should be noted that such a nucleosynthesis result that stems from
the low-$Y_e$ ($\sim 0.47-0.49$) matter in the early ejecta may be a
unique characteristics of collapsing O-Ne-Mg cores. In the present
model, the explosion sets in immediately after core bounce and the
ejecta rapidly expand, where the ejection of low $Y_e$ matter seems
unavoidable (see Fig.~1). For more massive progenitors ($> 10\,
M_\odot$), the explosion is expected to be more delayed and to eject
neutrino-processed matter, where the bulk of the ejecta may have $Y_e
\gtrsim 0.5$ \citep{Froh06, Bura06}.

As will be discussed in \S~4, the maximum overproduction factor of
$X/X_\odot = 5.7\times 10^5$ for model ST (Fig.~4) poses a severe
constraint on the occurrence of this type of events to be no more than
$1\%$ of all core-collapse supernovae. This is in agreement with the
conclusions by \citet{Hoff08}. In the simulation of \citet{Kita06}
\citep[also][]{Jank08a}, the deceleration of the shock in the outer
envelope slows the expansion of the ejecta only slightly. Therefore we
do not expect any substantial fallback of the once ejected matter
onto the remnant, as it is presumed to take place in the case of more 
massive progenitors \citep[e.g.,][]{Umed02}. In the following subsections
we thus explore possible modifications to model ST, which instead of 
fallback might provide a solution for moderating the extremely large
overproduction of, in particular, $^{90}$Zr.

\subsection{Uncertainty in the Equation of States}

First, we examine nucleosynthesis in the explosion model calculated
with the WH EoS in \citet{Kita06}. This model (hereafter WH) has
thermodynamic trajectories very similar to those of model ST, but a
slightly larger ejecta mass ($1.64\times 10^{-2}\, M_\odot$) and a
slightly lower minimal $Y_{e,\mathrm{i}}$ ($= 0.462$). Another EoS by
\citet{Shen98}, which is also currently available for core-collapse
simulations, falls between the ST and WH EoSs in terms of its
\textit{stiffness} and a variety of results of core-collapse
simulations, e.g.\ the radii of shock formation and stagnation or the
size of the neutrino luminosities \citep[see][]{Jank08b}.  Therefore,
we suspect that a comparison of the nucleosynthesis results between
models ST and WH well brackets the uncertainties arising from
different EoSs.

All nucleosynthesis calculations are repeated with the thermodynamic
trajectories and the initial compositions deduced from the
$Y_{e,\mathrm{i}}$-mass profile (Fig.~3) of model WH. Figure~7a shows
the mass fractions (\textit{bottom}) in the ejecta for models ST and
WH, along with their ratios $X_\mathrm{WH}/X_\mathrm{ST}$
(\textit{top}), as functions of the mass number $A$. Differences
exceeding a factor of 2 (indicated by \textit{dotted lines}) can be
seen for the light species with $A < 20$. However, the differences are
well below a factor of 2 for the dominant species in the vicinity of
the QSE peaks at $A\approx 4, 60$, and 90. We therefore conclude that
the uncertainties arising from different nuclear EoSs are not of great
importance, at least for the currently available versions of EoSs. We
note, however, that the WH EoS leads to a $60\%$ larger $^{56}$Ni mass
than that for the LS EoS (Table~1). This suggests that EoSs play a
crucial role to precisely determine the $^{56}$Ni ejecta mass from an
electron capture supernova.

\begin{figure}
\epsscale{1.0}
\plotone{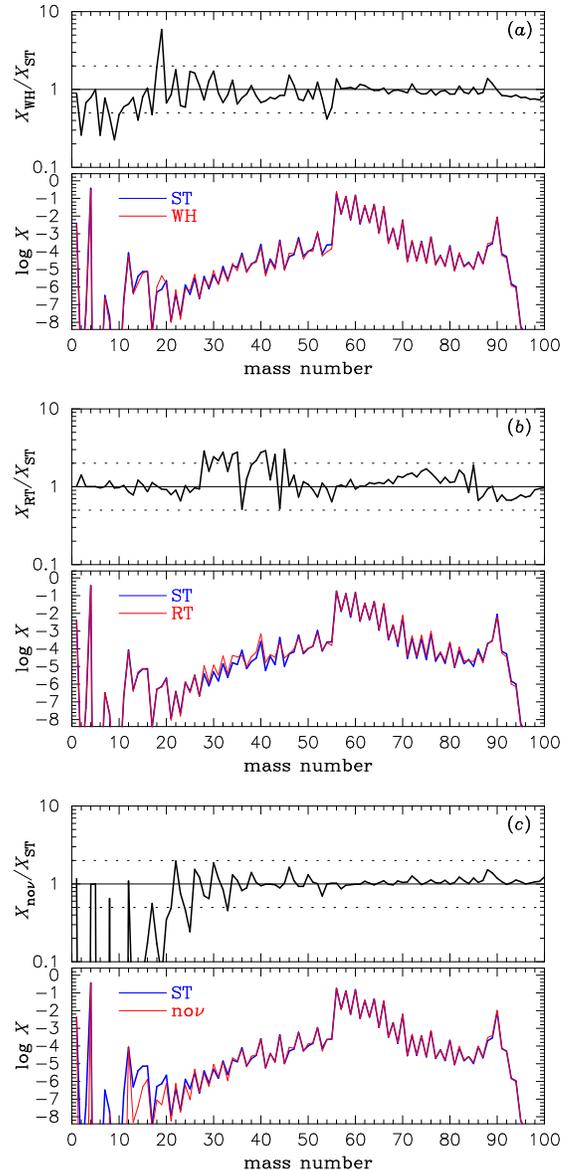}
\caption{Mass fractions (\textit{bottom} of each panel) in the ejecta
for models ST (\textit{blue}) and (\textit{a}) WH, (\textit{b}) RT,
(\textit{c}) no$\nu$ (\textit{red}; see the text), and their ratios
(\textit{top} of each panel), as functions of the mass number. In the
top panels, a factor of 2 difference is indicated by dotted lines.}
\end{figure}

\subsection{Uncertainty in the Nuclear Reaction Rates}

In all the present calculations, the nucleosynthetic flows proceed
along or in the vicinity of the $\beta$-stability line, forming the
QSE peaks at $A \approx 4$, 60, and 90. The nuclear masses of
relevance, which are the most important factors in this case, are
measured with high enough accuracy \citep{Audi03}. Therefore, we do
not expect a sizable change of the nucleosynthesis result arising from
the uncertainties in nuclear reaction rates.

As a test, we repeat all the nucleosynthesis calculations of model ST
with the theoretical rates \citep{Aika05} replaced by those of
\citet{Raus00}. Both data sets are based on the statistical
(Hauser-Feshbach) approach with the experimental masses whenever
available, but with different theoretical masses and different nuclear
level densities. The results (labelled by RT) are compared to those of
model ST in Figure~7b. Factors of $2-3$ difference can be seen for
the nuclei between $A = 27$ and 45 (Fig.~8, \textit{top}), which have,
however, very small mass fractions (\textit{bottom}). The differences
for dominant species around $A \approx 4$, 60, and 90 are well below a
factor of 2.


\subsection{Effects of Convection}

A potentially important effect that is lacking in the one-dimensional
simulations of \citet{Kita06} is convective mixing of the ejecta
\citep[see the results of a two-dimensional simulation of a collapsing
O-Ne-Mg core in][]{Jank08b}. Figure~8 shows the profiles of
temperature (\textit{top}), entropy (\textit{middle}), and $Y_e$
(\textit{bottom}) versus mass at early times ($t_\mathrm{pb} = 126,
150, 176, 200$, and 228~ms) and at the end of the simulation
($t_\mathrm{pb} = 810$~ms). We find a negative entropy gradient
forming at a temperature high enough for $\alpha$-processing ($T_9 >
3$, which is indicated by the dashed line in the top panel of
Fig.~9). This can cause convective overturn and one might speculate
that this could moderate the neutron-richness to some extent
\textit{before} the freezeout.

As a limiting case, we present the result with the initial composition
being determined by the $Y_{e,\mathrm{i}}$-mass profile shown by the
green line in Figure~3 (hereafter, model MX). We assume here that
the ejecta between 1.367 and $1.375\, M_\odot$ (indicated by vertical
dotted lines in Fig.~9) get completely mixed on microscopic scales
to have a mass-averaged constant
$Y_{e,\mathrm{i}}$ ($= 0.480$). We find that the species having the
largest overproduction, i.e., $^{90}$Zr in model ST, is replaced by
$^{64}$Zn with ten times smaller value in model MX (Fig.~9). The
largest overproduction of Zr (as element) is replaced by that of Zn
(Fig.~10). Because of the increased minimum value of 
$Y_{e,\mathrm{i}}$ ($Y_{e,\mathrm{min}} = 0.480$) the
overproduction of $^{90}$Zr becomes unimportant.

It should be noted, however, that a corresponding two-dimensional 
simulation (carried out until 262~ms after core bounce; 
see \citet{Jank08b} and M\"uller and Janka 2008, in preparation) 
with the same initial O-Ne-Mg core does not show any such small-scale
mixing. Instead, the accreted post-shock gas makes one quick overturn
(fully developed convection with many overturns does not occur), 
after which the rising material in
Rayleigh-Taylor mushrooms is directly ejected and self-similar 
expansion is quickly established. Hence, there may not be
sufficient time for mixing and homogenization on small scales. 
As will be discussed in \S~3.5, however, a slight increase of
$Y_{e,\mathrm{min}}$ (e.g., $\Delta Y_{e,\mathrm{i}} = 0.004$; 
from 0.464 to 0.468) is already enough to significantly moderate
the overproduction of $^{90}$Zr. Therefore, partial mixing of
the ejecta induced by convection might be sufficient to cure this
overproduction problem.

The negative entropy gradient may have a variety of other effects. One
is that the convective overturn could stretch the mean duration of the
neutrino irradiation of ejected matter
and could also lead to regions of ejecta with higher 
$Y_{e,\mathrm{i}}$. In fact, a recent
two-dimensional simulation with a more massive progenitor \citep[$15\,
M_\odot$,][]{Bura06} shows that the bulk of the ejecta turns out to be
proton-rich ($Y_e > 0.5$). In the present case, however, the
convective overturn may not have exactly the same effects as in 
more massive progenitors, because the steep density gradient of
the O-Ne-Mg core with its transition to an extremely dilute 
H-rich envelope makes the core structure distinctively different 
from that of more massive stars. As a consequence, the supernova
ejecta accelerate much faster than in more massive stars, and the 
convective pattern in the overturn region freezes out in the 
self-similar expansion more quickly. Therefore convective mixing
is rather inefficient and, moreover, it is possible that 
the neutrino-heated bubbles rise so rapidly away from the
neutrino-sphere that clumps of low-$Y_e$ material get ejected, a 
possibility that may be anticipated from the bottom panel of
Figure~8 (e.g., the red and green lines there). This might increase the
overproduction of other $N=50$ species ($^{86}$Kr, $^{87}$Rb,
$^{88}$Sr, $^{89}$Y). In fact, some neutron-rich blobs (down to $Y_e
\approx 0.41$, but with a tiny mass) are found in the preliminary
results of the mentioned two-dimensional simulation \citep[M\"uller
and Janka 2008, in preparation, see also][]{Jank08b}.

It should be noted that even two-dimensional models with their
constraint of axisymmetry may not yield
a sufficiently accurate mass distribution of the ejecta as a function
of $Y_e$. Such a sensitive information might ultimately require 
three-dimensional simulations to allow for reliable conclusions 
concerning nucleosynthesis yields and production factors.

\begin{figure}
\epsscale{1.0}
\plotone{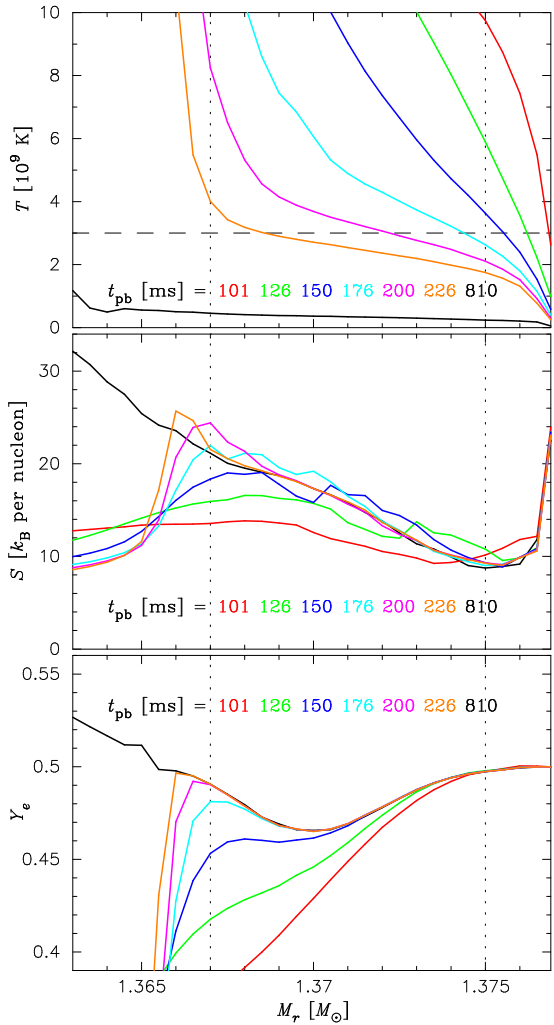}
\caption{Profiles of temperature (\textit{upper}), entropy
(\textit{middle}), and $Y_e$ (\textit{bottom}) versus enclosed mass
$M_r$ at early times ($t_\mathrm{pb} = 100-230$~ms) and at the end of
simulation ($t_\mathrm{pb} = 810$~ms). See the text for the meaning of 
the vertical dotted lines and the horizontal dashed line.}
\end{figure}

\begin{figure}
\epsscale{1.0}
\plotone{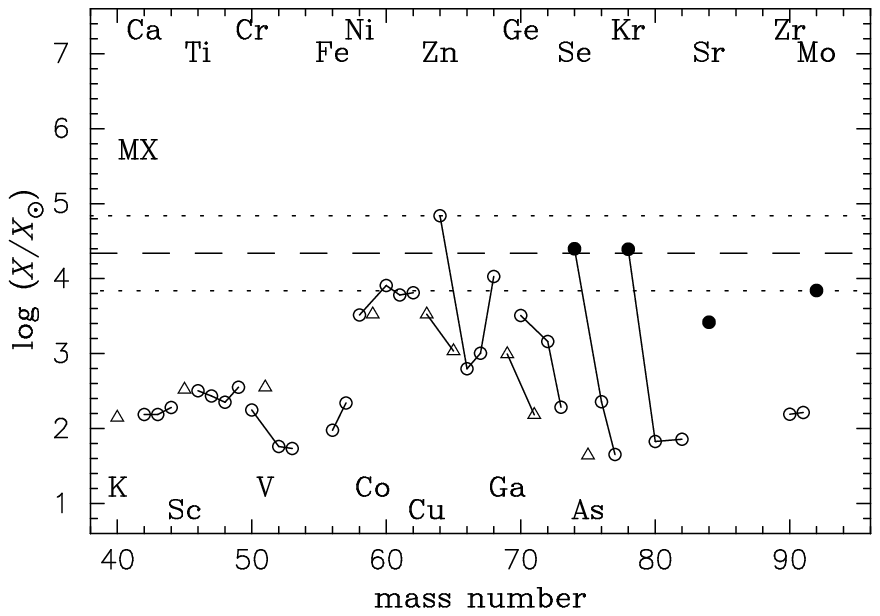}
\caption{Same as Figure~4, but for model MX.}
\end{figure}

\begin{figure}
\epsscale{1.0}
\plotone{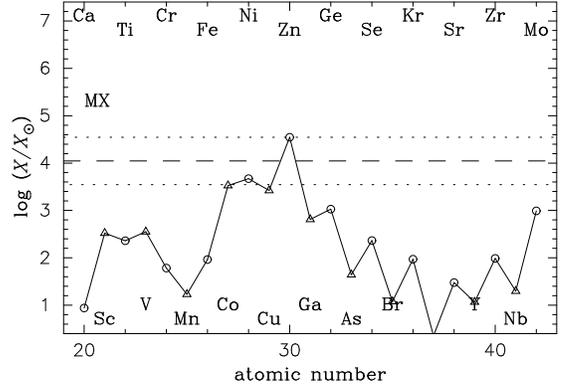}
\caption{Same as Figure~5, but for model MX.}
\end{figure}

\subsection{Small $Y_e$ Variation}

 From Figure~6 we conclude that the large overproduction of $^{90}$Zr
in model ST is mainly due to the neutron-rich ejecta with
$Y_{e,\mathrm{i}}$ between $Y_{e,\mathrm{min}} = 0.464$ and 0.470. The
demonstration in \S~3.4 (Fig.~9) shows that boosting
$Y_{e,\mathrm{min}}$ to 0.480 indeed removes the overproduction of
$^{90}$Zr. Motivated by this result, we explore in this subsection
that how much increase of $Y_{e,\mathrm{min}}$ could cure this extreme
overproduction problem.  We repeat that the $Y_{e,\mathrm{i}}$ values
are obtained from state-of-the-art simulations with sophisticated,
energy-dependent neutrino transport in \citet{Kita06}.  Nevertheless
it is a great challenge to determine $Y_e$ to an accuracy of a few
percent, because the neutron-to-proton ratio in the ejecta is
established by a delicate competition of electron neutrino and
antineutrino captures and their inverse reactions, in particular
around the radius where the kinetic equilibrium of these processes
breaks down because the rates become slower than the expansion rate of
the accelerating ejecta \citep[for a detailed discussion,
see][]{Froh06}. Besides the limited numerical resolution (in
particular of the neutrino energy spectra) a variety of other effects
can be imagined to imply uncertainties at the percent level, for
example the potential effects of convective mixing on microscopic
scales (\S~3.4), future refinements in the employed microphysics
(e.g., EoSs, neutrino interaction rates, electron captures rates),
possible effects due to nonstandard neutrino properties (e.g., flavor
oscillations in the supernova core), and the uncertainties associated
with the stellar evolution calculations of $8-10\, M_\odot$ stars.

In order to test small variations of $Y_{e,\mathrm{min}}$, we examine
nucleosynthesis for the trajectories of model ST but with the original
$Y_{e,\mathrm{i}}$ profile replaced by $Y_{e,\mathrm{i}} +
(0.500-Y_{e,\mathrm{i}}) \times f$. The multiplicative factor $f$ is
taken to be 0.1, 0.2, and 0.3 (hereafter, models FP1, FP2, and
FP3). The $Y_{e,\mathrm{i}}$-$M_r$ profiles of these models are similar,
but each model has a different $Y_{e,\mathrm{min}}$
(see Fig.~3 for model FP3). The
values of $Y_{e,\mathrm{min}}$ for these models are slightly increased
compared to that of model ST to be 0.468, 0.471, and 0.475,
respectively (Table~1). 

We find that the production of $^{90}$Zr is extremely sensitive to
$Y_{e,\mathrm{min}}$ (Figure~11).  The increase of
$Y_{e,\mathrm{min}}$ by only $1-2\%$ ($\Delta Y_{e,\mathrm{i}} =
0.004$ and 0.011 in models FP1 and FP3, respectively) reduces the
production factor of $^{90}$Zr by roughly two orders of
magnitude. Other species like $^{64}$Zn, $^{74}$Se, and $^{78}$Kr then
possess the largest production factors.  In the case of model FP3
these have one tenth of the values in model ST. Figure~12 indicates
that electron capture supernovae could be the dominant sources of the
elemental abundance of Zn, if such a slightly higher
$Y_{e,\mathrm{i}}$ was correct.

\begin{figure}
\epsscale{1.0}
\plotone{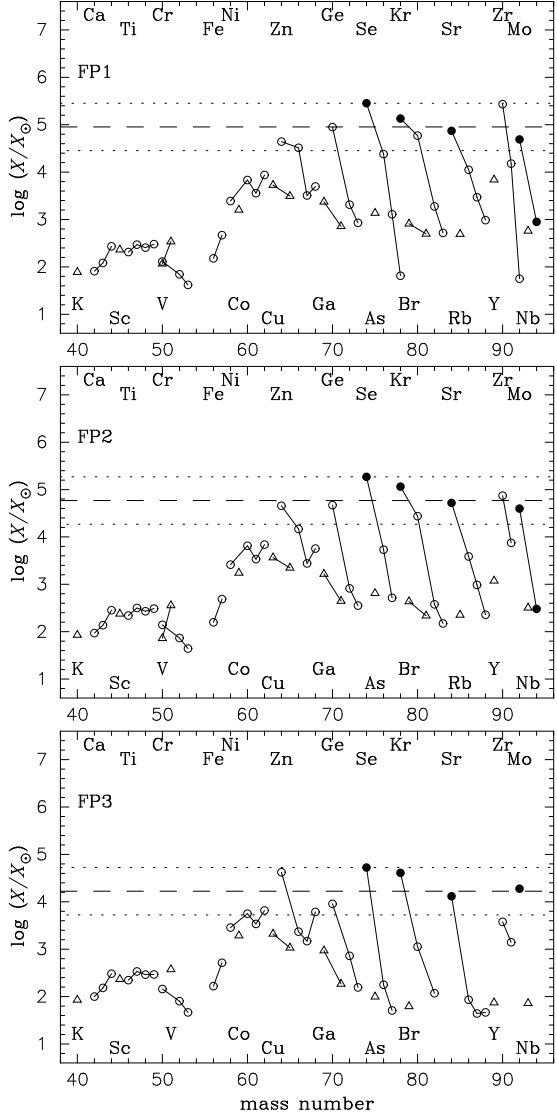}
\caption{Same as Figure~4, but for models FP1 (\textit{top}), FP2
(\textit{middle}), and FP3 (\textit{bottom}).}
\end{figure}

\begin{figure}
\epsscale{1.0}
\plotone{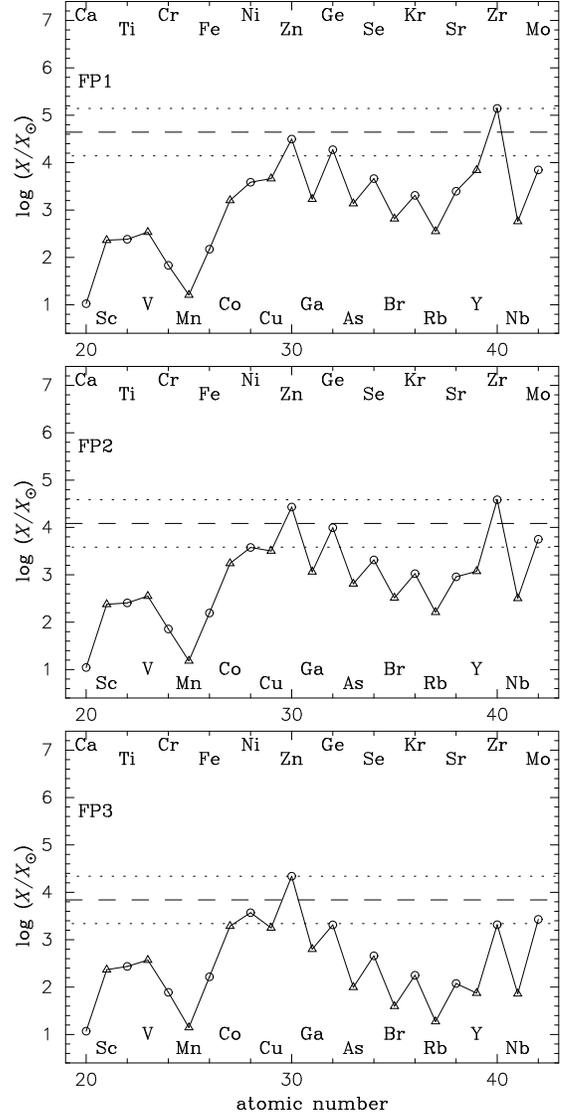}
\caption{Same as Figure~5, but for models FP1 (\textit{top}), FP2
(\textit{middle}), and FP3 (\textit{bottom}).}
\end{figure}

The results for $f = -0.1$, $-0.2$, and $-0.3$ (hereafter, models FM1,
FM2, and FM3) are shown in Figures~13 and 14. The values of
$Y_{e,\mathrm{min}}$ are then decreased to 0.460, 0.457, and 0.453,
respectively (Table~1). The $Y_{e,\mathrm{i}}$-$M_r$ profile for model
FM3 is also displayed in Figure~3.  The overproduction of $^{90}$Zr
becomes more serious in these models, and other $N=50$ species
($^{88}$Sr and $^{89}$Y) enter into the normalization band in the case
of model FM3.  The results of \citet[][Fig.~1]{Hoff08} resemble those
for our model FM3 rather than those for model ST. This is probably
explained by the fact that \citet{Hoff08} perform their evaluation for
ejecta with a value of $Y_{e,\mathrm{min}}= 0.454$, which is close to
$Y_{e,\mathrm{min}}$ in model FM3 ($=0.453$), but slightly smaller
than $Y_{e,\mathrm{min}}$ in model ST. The small difference between
the ejecta conditions in \citet{Hoff08} and those of model ST
originates from a different density structure assumed for the dilute
H/He envelope around the collapsing O-Ne-Mg core considered by
\citet{Hoff08} \citep[see][for more detail]{Jank08a}.

The results presented in this subsection imply that the overproduction
of $^{90}$Zr may not be as serious as reported by \citet{Hoff08},
because an increase of $Y_{e,\mathrm{min}}$ by only $2\%$ cures this
problem. This is a significant improvement compared to the situation 
in the older simulations by
\citet{Mayl88}, where $Y_{e,\mathrm{min}}$ was found to be around 0.4
and a substantial change ($\sim 20\%$) of $Y_{e,\mathrm{min}}$ was
necessary to avoid the overproduction of $N = 50$ nuclei.

\begin{figure}
\epsscale{1.0}
\plotone{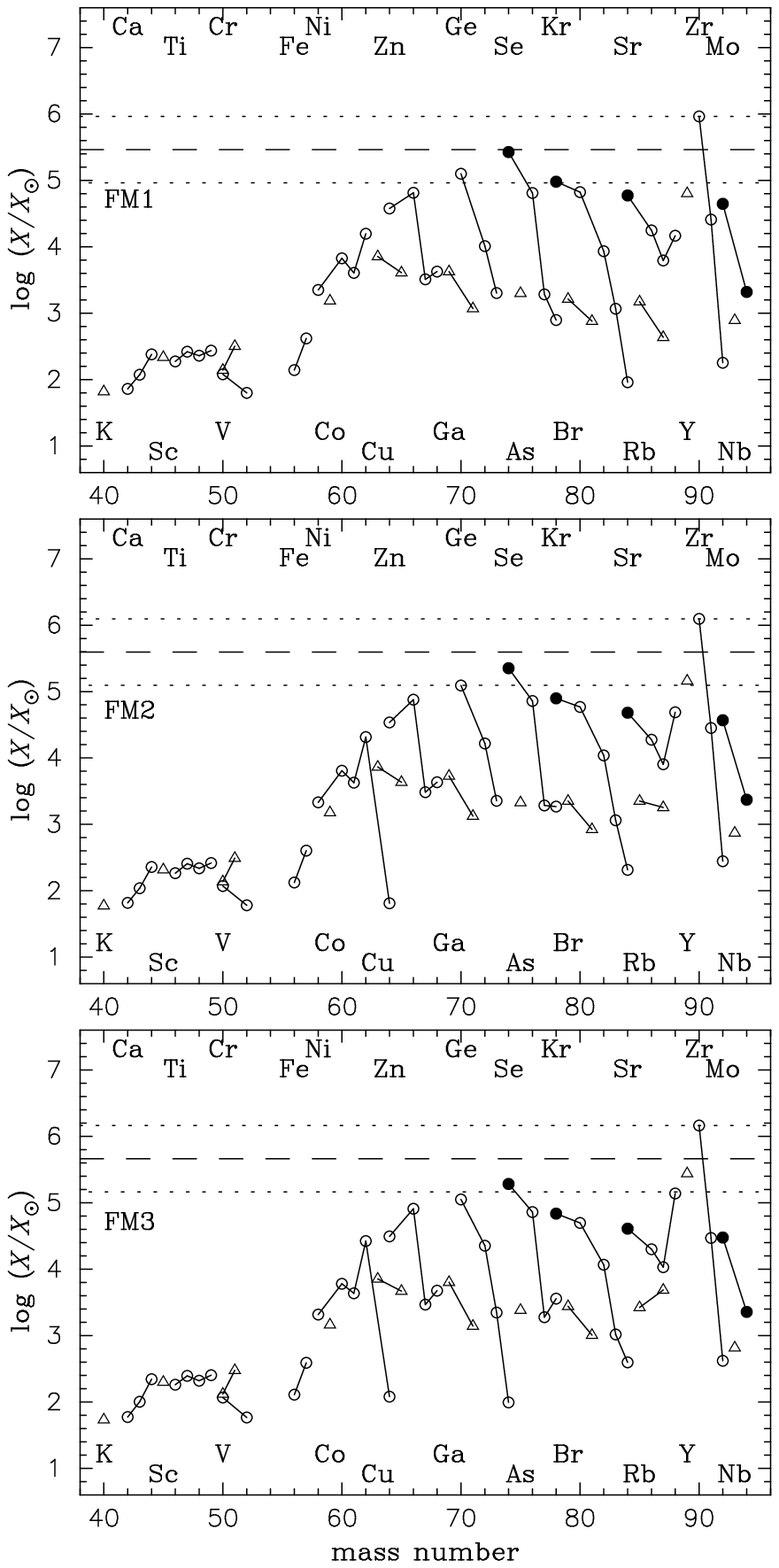}
\caption{Same as Figure~4, but for models FM1 (\textit{top}), FM2
(\textit{middle}), and FM3 (\textit{bottom}).}
\end{figure}

\begin{figure}
\epsscale{1.0}
\plotone{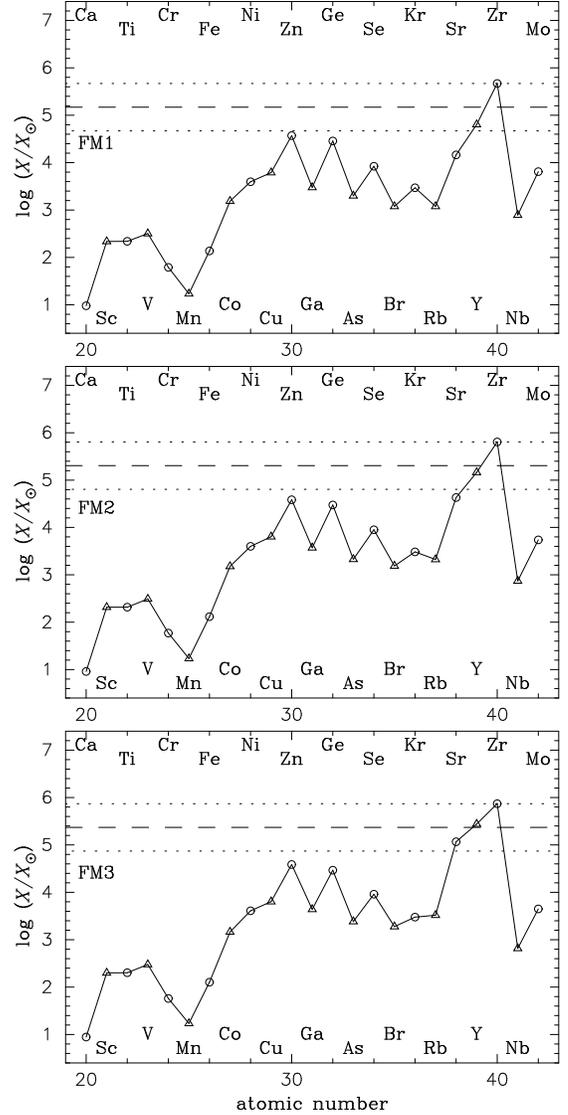}
\caption{Same as Figure~5, but for models FM1 (\textit{top}), FM2
(\textit{middle}), and FM3 (\textit{bottom}).}
\end{figure}

\subsection{$\nu$p-process}

The $\nu$p-process (or neutrino-induced rp-process) is now believed to
be a promising nuclear process that synthesizes light $p$-nuclei up to
$A\sim 110$ \citep{Froh06, Prue06, Wana06}. In this process, a
fraction of free protons are converted to neutrons by neutrino capture
in the early proton-rich supernova ejecta. The $\beta$-waiting points
on the classical rp-process path (e.g., $^{64}$Ge with the half-life
of 1.06 minutes) are then bypassed via much faster neutron
capture. The current models contain proton-rich ejecta (up to
0.53, Figs.~1 and~3), where one may expect the occurrence of the
$\nu$p-process.

The $\nu$p-process plays, however, no role in producing the
$p$-nuclei, although all the examined models in this study include
neutrino-induced reactions on free nucleons and $\alpha$ particles. In
order to test the effect of neutrinos, the result without
neutrino-induced reactions (labelled no$\nu$, otherwise using the same
input as in model ST) is compared with that of model ST (Fig.~7c). We
find that neutron capture, which is absent without neutrino-induced
reactions, diminishes the nuclei with $A \le 20$. The mass fractions
of these isotopes are, however, very small compared to the dominant
species at $A \approx 4$, 60, and 90. No substantial differences can
be seen for the nuclei with $A > 50$ because of the moderate
proton-richness (up to $Y_{e,\mathrm{i}} = 0.53$), moderate entropy
(up to $30\,k_\mathrm{B}$ per nucleon, where $k_\mathrm{B}$ is the
Boltzmann constant), and the fast expansion of the ejecta. The
proton-rich ejecta quickly cool down below $T_9 \approx 2-3$, which is
the relevant temperature range for the $\nu$p-process to take place
(i.e., the proton captures are fast and their inverse reactions
slow). The $\nu$p-process might be efficient only in core-collapse
supernovae from more massive progenitors (e.g., $\ga 15\, M_\odot$),
which have a shallower density gradient at the core edge and a denser
envelope and thus their explosions develop in a different way with
different conditions for nucleosynthesis.


\section{Contribution to Galactic Chemical Evolution}

We now discuss a possible contribution of electron capture supernovae
to Galactic chemical evolution. First, we suppose that model ST, which
has the largest overproduction factor for $^{90}$Zr (Fig.~4), is
representative of this type of supernovae. Let us assume that electron
capture supernovae produce all $^{90}$Zr in nature, and the other
supernovae from the progenitors more massive than $10\, M_\odot$
produce $^{16}$O with a typical amount of
$M_\mathrm{other}(^{16}\mathrm{O}) = 1.5\, M_\odot$ per event. Here,
$M_\mathrm{other}(^{16}\mathrm{O})$ is taken to be the
initial-mass-function averaged yield of \citet{Nomo06} between 13 and
$40\, M_\odot$ (solar metallicity models). The contribution of
electron capture supernovae to the Galactic $^{16}$O is negligible
(Table~2). If we assume the number fraction of electron capture
supernovae relative to all core-collapse supernovae to be $f_*$, we
have the relation
\begin{equation}
\frac{f_*}{1-f_*} 
= \frac{X(^{90}\mathrm{Zr})_\odot/X(^{16}\mathrm{O})_\odot}
       {M(^{90}\mathrm{Zr})/M_\mathrm{other}(^{16}\mathrm{O})}
= 0.029,
\end{equation}
where $X(^{16}\mathrm{O})_\odot = 6.6 \times 10^{-3}$,
$X(^{90}\mathrm{Zr})_\odot = 1.5 \times 10^{-8}$ \citep{Lodd03}, and
$M(^{90}\mathrm{Zr}) = 1.2 \times 10^{-4}\,M_\odot$ (Table~1).
Therefore we expect the frequency of electron capture supernovae to be
no more than 1\% of all core-collapse events, when taking into account
that there are also other sources of $^{90}$Zr \citep[$81\%$ is from
the $s$-process,][]{Burr00}. As discussed in \S~3.5, however, the
production of $^{90}$Zr is extremely sensitive to $Y_{e,\mathrm{i}}$,
changing more than 2 orders of magnitude with only 2\% variation of
$Y_{e,\mathrm{i}}$ (Table~1). Thus, we do not consider the
overproduction of $^{90}$Zr to give a tight constraint on the
occurrence of electron capture supernovae.

Instead, we propose the abundance of $^{64}$Zn, whose production is
insensitive to small variations of $Y_{e,\mathrm{i}}$ (Table~1), to
serve as a strong constraint on the occurrence of this type of
supernovae. As an example, we consider model FP3 (Fig.~11;
\textit{bottom}) as representative of electron capture supernovae, in
which the largest overproduction is shared by $^{64}$Zn, $^{74}$Se,
and $^{78}$Kr. The origins of these isotopes have not been well
identified, although the latter two $p$-isotopes might be produced to
some extent by the $\gamma$-process \citep{Raye95} or $\nu$p-process
\citep{Froh06, Prue06, Wana06} in core-collapse supernovae. We
therefore can assume that all $^{64}$Zn (and $^{74}$Se, $^{78}$Kr) in
nature is produced by electron capture supernovae. In this case,
equation~(1) with $^{90}$Zr replaced by $^{64}$Zn gives $f_* = 0.28$,
where $X(^{64}\mathrm{Zn})_\odot = 1.1 \times 10^{-6}$ \citep{Lodd03}
and $M(^{64}\mathrm{Zn}) = 6.5 \times 10^{-4}\,M_\odot$ (Table~1). We
thus conclude that the upper limit of the frequency of electron
capture supernovae is around $30\%$ of all the core-collapse
events. This is in good agreement with the upper limit estimated from
a recent study of stellar evolution \citep[$\sim 20\%$,][]{Poel08} as
well as the estimate for the rate of SN~2008S-like transients
\citep[$\sim 20-30\%$,][]{Thom08}. Our result implies that a
significant fraction of $8-10\, M_\odot$ is allowed to enter into the
supernova channel from the nucleosynthetic point of view.

For a practical use, the yields of all stable (Table~2) and some
unstable (Table~3) isotopes for models ST and FP3 are presented. For
the study of Galactic chemical evolution, one can use those of model
FP3 as representative of electron capture supernovae, keeping in mind that
the $Y_{e,\mathrm{i}}$ profile (i.e., the initial composition) of
this model was slightly modified. Nevertheless, it would be interesting
to see if this type of supernovae was the dominant source of Zn and
of the most mysterious $p$-nucleus $^{92}$Mo. In fact, all the
previous models of supernova nucleosynthesis (with typical explosion
energy), except for \textit{hypernovae} \citep{Umed02}, have failed to
explain the solar inventory of $^{64}$Zn (which is the most abundant
isotope of this element). We do not expect a substantial contribution
to the abundance of $^{64}$Zn from the later ($t_\mathrm{pb} > 1$~s)
neutrino-driven wind \citep{Hoff96, Wana06}, whose total mass is no
more than $0.001\, M_\odot$ with $^4$He being the dominant species
\citep[e.g.,][]{Wana07}. The $p$-isotope $^{92}$Mo can be synthesized
by the $\nu p$-process to some extent, but still falls a few times
short of what is expected from the neighboring $p$-nuclei
\citep[$^{84}$Sr, $^{94}$Mo, and $^{96,98}$Ru,][]{Prue06, Wana06}.

\section{Origin of Faint Supernovae?}

Our result confirms that electron capture supernovae produce very
little $^{56}$Ni ($\approx 0.002-0.004\, M_\odot$, Table~1) compared
to $\sim 0.1\, M_\odot$ in the case of more massive progenitors
\citep[e.g.,][]{Nomo06}. This is a consequence of the small ejecta
mass ($= 0.0139\, M_\odot$ without the H-rich envelope) and also of
the neutron-richness of the bulk of the ejecta ($Y_{e,\mathrm{i}} <
0.49$ for mass shells in the range of $1.367-1.373\, M_\odot$,
Fig.~6), because of which $^{56}$Ni is not a dominant species to be
produced. The yield of $^{56}$Ni, which is mainly created in matter
with $Y_{e,\mathrm{i}} \sim 0.5$, is insensitive to a small variation
of $Y_e$ (Table~1). We also do not expect a significant contribution
from the later ($t_\mathrm{pb} > 1$~s) neutrino wind. The uncertainty
arising from the nuclear EoS seems larger than that from $Y_e$, being
about a factor of 2 (Table~1) for the currently available versions of
EoSs. The uncertainty might become larger if a variety of EoSs are
available in the future. The convective motion near the mass cut may
also affect the $^{56}$Ni mass (\S~3.4), which needs multi-dimensional
simulations.

The expected small amount of $^{56}$Ni as well as the low explosion
energy of electron capture supernovae have been proposed as an
explanation of the observed properties of faint SNe~II-P \citep[e.g.,
SN~1997D,][]{Chug00, Kita06} and of the low luminosity of
SN~2008S-like transients \citep[possibly a new sub-class of
SNe~IIn,][]{Prie08, Thom08}. The former ones, namely SNe 1994N, 1997D,
1999br, 1999eu, 2001dc, and 2005cs, have been observationally defined
as the class of low-luminosity Ni-poor SNe~II-P \citep[][]{Past04,
Past06}, whose incidence is estimated to be as high as $4-5\%$ of all
SNe~II.  The estimated $^{56}$Ni masses of $\sim 0.002-0.008\,
M_\odot$ for these low-luminosity SNe~II-P \citep{Zamp03, Past04,
Past06, Hend05} are in reasonable agreement with the present result
from electron capture supernovae. An alternative possibility is the
origin of such supernovae from much more massive stars ($\gtrsim 20\,
M_\odot$) with low explosion energies, which suffer from fallback of
freshly synthesized $^{56}$Ni \citep[e.g.,][]{Tura98, Bene01, Nomo03,
Zamp03}. This is due to the inferred large envelope masses for the
low-luminosity SNe~II-P \citep[$\sim 14-40\, M_\odot$,][]{Hend05} that
favor massive progenitors. A recent analysis of the progenitors of
SNe~II-P by \citet{Smar08} indicates, however, that low-luminosity
supernovae with low $^{56}$Ni production are likely to arise from low
mass progenitors near the minimum mass limit for core-collapse
supernovae. A lack of $\alpha$-elements such as O and Mg in the case
of electron capture supernovae will be a key to spectroscopically
distinguish between these two scenarios \citep{Kita06}.

The Crab nebula (the relic of SN~1054) is known to have a low kinetic
energy \citep[$\sim 4\times 10^{49}$~erg,][]{Chev85} and a small
amount of $\alpha$-elements \citep{Davi82}. An electron capture
supernova has been suggested to be the origin of the Crab nebula
\citep{Nomo82, Nomo85, Kita06}. Our result with the little production
of $\alpha$-elements and iron supports this idea. The Ni/Fe ratios
($\approx 1-2$, except for the extreme model MX, Table~1), which are
at least 20 times larger than the solar value ($=0.058$) can be
considered as an additional support, because they are in reasonable
agreement with the reported high Ni/Fe ratio of the Crab nebula
\citep[$\sim 10$ times solar;][]{Henr84, Hudg90}.

Recently, \citet{Maca08} investigated gaseous regions of the Crab
nebula and inferred from their photoionization calculations that the
abundance of a large component of the nebula appears to be He-rich and
$\textrm{C}/\textrm{O} > 1$. According to the $8-10\, M_\odot$ star
models \citep{Nomo82, Nomo84}, the He-rich envelope for $M \lesssim
9.5\, M_\odot$ has $\textrm{C}/\textrm{O} < 1$ because of the
preceding CNO-cycle, while that for $M \gtrsim 9.5\, M_\odot$ has
$\textrm{C}/\textrm{O} > 1$ owing to the 3$\alpha$-reactions. The
$9.6\, M_\odot$ star in \citet[][case~2.4]{Nomo84}, for example, has
carbon and oxygen mass fractions of 0.022 and 0.0033 in the He-burning
convective layer \citep[corresponding solar values are 0.0025 and
0.0066, respectively;][]{Lodd03}. This enhanced carbon abundance (ten
times that of the solar value) is consistent with that reported by
\citet[][]{Maca08}. It should be noted that stars with initial masses
of $\sim 9.5-10\, M_\odot$ have almost identical core structures to
our considered star of $8.8\, M_\odot$, except for the outermost
oxygen mass fraction \citep{Nomo84, Nomo87}. Therefore, the explosion
of a star with an initial mass of $\sim 9.5-10\, M_\odot$ and with an
O-Ne-Mg core can be the origin of the Crab remnant.

We also note that the dredge-up of the material from the He-layer into
the H-layer enhances carbon in the envelope of the AGB star
\citep{Nomo87}. This would be more efficient in enhancing carbon than
He-thermal pulses, and dust could be easily formed to induce mass
loss. This may result in a deeply dust-enshrouded object such as the
progenitor of SN~2008S \citep{Prie08, Thom08}. For the $9.6\, M_\odot$
star in \citet[][case~2.4]{Nomo84}, the duration of the AGB phase is
estimated to be $4\times 10^4$~yr (that becomes shorter for a more
massive case), which is in reasonable agreement with the inferred
dust-enshrouded phase for SN~2008S-like transients \citep[$\lesssim
10^4$~yr,][]{Thom08}. This might also imply the mass range of the
stars that end their lives as electron capture supernovae to be $\sim
9.5-10\, M_\odot$, whose frequency, $\sim 7-8\%$ of all the
core-collapse events, satisfies the constraint from our
nucleosynthesis results ($< 30\%$; \S~4).

\section{Conclusions}

We have investigated the nucleosynthesis during the first 810~ms after
core bounce in an explosion from a collapsing star with O-Ne-Mg core
(electron capture supernova) and an initial mass of $8.8\,
M_\odot$. The thermodynamic trajectories are taken from the
self-consistent explosion models of \citet{Kita06}, which were
computed with the initial stellar model of \citet{Nomo84, Nomo87}. Our
main conclusions can be summarized as follows.

1. Our unmodified model (ST) results in (i) little production of
$\alpha$-elements and iron, (ii) large production of $^{64}$Zn,
$^{70}$Ge, and in particular, $^{90}$Zr, and (iii) production of some
light $p$-nuclei ($^{74}$Se, $^{78}$Kr, $^{84}$Sr, and
$^{92}$Mo). This is a consequence of the ejection of a sizable amount
of neutron-rich matter ($6 \times 10^{-3}\, M_\odot$ with $Y_e =
0.46-0.49$). If we assume this model to be representative of electron
capture supernovae, the occurrence of this type of supernovae is
limited to be no more than $1\%$ of all core-collapse events. We do
not think, however, that the production of $^{90}$Zn serves as a
strong constraint, because it is easily affected by a small variation
of $Y_e$. The $\nu p$-process does not play any role for the
production of $p$-nuclei in the present supernova model.

2. The uncertainties in the nuclear EoS and the nuclear reaction rates
do not substantially affect the nucleosynthesis results. In contrast,
the effects of convection, which are not included in the
one-dimensional simulations of \citet{Kita06}, are expected to be
large and may change the initial $Y_e$ distribution to some
extent. The overproduction of $^{90}$Zr is moderated if the minimum
$Y_e$ is only $1-2\%$ larger than that in the unmodified model ST. In
this case (our model FP3) the largest overproduction, which is
observed for $^{64}$Zn, $^{74}$Se, and $^{78}$Kr, is reduced to one
tenth of that of the unmodified model. The robustness of the $^{64}$Zn
production against small variations of $Y_e$ provides an upper limit
to the occurrence of electron capture supernovae to be about $30\%$ of
all stellar core-collapse events. Electron capture supernovae can be
significant contributors to the Galactic inventories of $^{64}$Zn (the
most abundant isotope of Zn) and some light $p$-nuclei (e.g.,
$^{92}$Mo), if the assumed slightly larger values of $Y_e$ were
correct.

3. The high Ni/Fe ratio ($= 1-2$) and the small production of
$\alpha$-elements, as well as the low explosion energy \citep[$1-2
  \times 10^{50}$~erg,][]{Kita06, Jank08a,Jank08b}, support the
hypothesis that the Crab nebula is the remnant of
an electron capture supernova \citep{Nomo82, Davi82}.

4. SN~2008S-like transients, whose progenitors are deeply
dust-enshrouded massive stars, are likely to be electron capture
supernovae of AGB stars. This implies that electron capture supernovae
constitute a newly identified sub-class of SNe~IIn \citep{Prie08,
  Thom08}.

5. The ejecta mass of $^{56}$Ni is $0.002-0.004\, M_\odot$, which is
in reasonable agreement with estimates for observed low-luminosity
supernovae. The amount might be, however, affected by the convective
motion near the mass cut. Multi-dimensional studies will clarify the
effect of convection on the production of $^{56}$Ni as well as
$^{90}$Zr.




\acknowledgements

We are grateful to an anonymous referee for important comments. This
research has been supported in part by World Premier International
Research Center Initiative (WPI Initiative), MEXT, Japan, and by the
Grant-in-Aid for Scientific Research of the JSPS (17740108, 18104003,
18540231, 20540226) and MEXT (19047004, 20040004).  In Garching, the
project was supported by the Deutsche Forschungsgemeinschaft through
the Transregional Collaborative Research Centers SFB/TR~27 ``Neutrinos
and Beyond'' and SFB/TR~7 ``Gravitational Wave Astronomy'', and the
Cluster of Excellence EXC~153 ``Origin and Structure of the Universe''
({\tt http://www.universe-cluster.de}). The computations were done at
the Rechenzentrum Garching and at the High Performance Computing
Center Stuttgart (HLRS) under grant number SuperN/12758.

\begin{deluxetable}{cccccc}
\tablecaption{Yields in Units of Solar Masses}
\tablewidth{0pt}
\tablehead{
\colhead{Model} &
\colhead{$Y_{e,\mathrm{min}}$} &
\colhead{$^{56}\mathrm{Ni}$} &
\colhead{$^{64}$Zn} &
\colhead{$^{90}$Zr} &
\colhead{Ni/Fe} 
}
\startdata
ST  & 0.464 & 2.50E$-$03 & 6.38E$-$04 & 1.21E$-$04 & 1.65 \\
WH  & 0.462 & 4.06E$-$03 & 7.31E$-$04 & 1.39E$-$04 & 1.27 \\
RT  & 0.464 & 2.52E$-$03 & 6.94E$-$04 & 7.83E$-$05 & 1.58 \\
MX  & 0.480 & 1.67E$-$03 & 1.07E$-$03 & 3.32E$-$08 & 3.01 \\
FP1 & 0.468 & 2.62E$-$03 & 6.83E$-$04 & 5.75E$-$05 & 1.55 \\
FP2 & 0.471 & 2.76E$-$03 & 7.08E$-$04 & 1.59E$-$05 & 1.46 \\
FP3 & 0.475 & 2.91E$-$03 & 6.51E$-$04 & 8.04E$-$07 & 1.36 \\
FM1 & 0.460 & 2.41E$-$03 & 5.83E$-$04 & 1.96E$-$04 & 1.73 \\
FM2 & 0.457 & 2.32E$-$03 & 5.31E$-$04 & 2.66E$-$04 & 1.82 \\
FM3 & 0.453 & 2.24E$-$03 & 4.83E$-$04 & 3.11E$-$04 & 1.92  
\enddata
\end{deluxetable}

\begin{deluxetable}{cccccc}
\tablecaption{Yields of Stable Isotopes (in Units of $M_\odot$)}
\tablewidth{0pt}
\tablehead{
\colhead{Species} &
\colhead{ST} &
\colhead{FP3} &
\colhead{Species} &
\colhead{ST} &
\colhead{FP3}
}
\startdata
$^{  1}$H \dotfill & 5.55E$-$05 & 3.84E$-$05 & $^{ 58}$Ni\dotfill & 1.79E$-$03 & 2.21E$-$03 \\
$^{  2}$H \dotfill & 1.98E$-$13 & 1.35E$-$13 & $^{ 60}$Ni\dotfill & 2.09E$-$03 & 1.72E$-$03 \\
$^{  3}$He\dotfill & 6.52E$-$10 & 6.49E$-$10 & $^{ 61}$Ni\dotfill & 5.04E$-$05 & 4.62E$-$05 \\
$^{  4}$He\dotfill & 5.12E$-$03 & 5.55E$-$03 & $^{ 62}$Ni\dotfill & 5.11E$-$04 & 2.88E$-$04 \\
$^{  6}$Li\dotfill & 6.70E$-$15 & 6.57E$-$15 & $^{ 64}$Ni\dotfill & 2.55E$-$07 & 5.38E$-$09 \\
$^{  7}$Li\dotfill & 4.52E$-$09 & 5.31E$-$09 & $^{ 63}$Cu\dotfill & 5.96E$-$05 & 1.97E$-$05 \\
$^{  9}$Be\dotfill & 2.05E$-$14 & 1.99E$-$14 & $^{ 65}$Cu\dotfill & 1.56E$-$05 & 4.59E$-$06 \\
$^{ 10}$B \dotfill & 2.72E$-$14 & 2.45E$-$14 & $^{ 64}$Zn\dotfill & 6.38E$-$04 & 6.51E$-$04 \\
$^{ 11}$B \dotfill & 3.02E$-$09 & 3.34E$-$09 & $^{ 66}$Zn\dotfill & 4.54E$-$04 & 2.16E$-$05 \\
$^{ 12}$C \dotfill & 1.18E$-$06 & 1.48E$-$06 & $^{ 67}$Zn\dotfill & 4.44E$-$06 & 2.03E$-$06 \\
$^{ 13}$C \dotfill & 6.82E$-$09 & 7.09E$-$09 & $^{ 68}$Zn\dotfill & 2.89E$-$05 & 3.90E$-$05 \\
$^{ 14}$N \dotfill & 5.54E$-$08 & 4.08E$-$08 & $^{ 70}$Zn\dotfill & 9.72E$-$12 & 2.14E$-$14 \\
$^{ 15}$N \dotfill & 1.02E$-$07 & 8.25E$-$08 & $^{ 69}$Ga\dotfill & 1.93E$-$06 & 5.68E$-$07 \\
$^{ 16}$O \dotfill & 1.03E$-$07 & 1.13E$-$07 & $^{ 71}$Ga\dotfill & 3.96E$-$07 & 7.64E$-$08 \\
$^{ 17}$O \dotfill & 5.17E$-$11 & 5.33E$-$11 & $^{ 70}$Ge\dotfill & 8.29E$-$05 & 6.61E$-$06 \\
$^{ 18}$O \dotfill & 7.16E$-$09 & 5.41E$-$09 & $^{ 72}$Ge\dotfill & 5.05E$-$06 & 7.10E$-$07 \\
$^{ 19}$F \dotfill & 1.02E$-$08 & 8.95E$-$09 & $^{ 73}$Ge\dotfill & 4.11E$-$07 & 4.32E$-$08 \\
$^{ 20}$Ne\dotfill & 3.12E$-$08 & 4.08E$-$08 & $^{ 74}$Ge\dotfill & 5.16E$-$09 & 6.86E$-$12 \\
$^{ 21}$Ne\dotfill & 1.63E$-$10 & 2.00E$-$10 & $^{ 76}$Ge\dotfill & 2.91E$-$14 & 4.90E$-$20 \\
$^{ 22}$Ne\dotfill & 5.37E$-$09 & 5.82E$-$09 & $^{ 75}$As\dotfill & 3.35E$-$07 & 1.84E$-$08 \\
$^{ 23}$Na\dotfill & 3.33E$-$10 & 4.46E$-$10 & $^{ 74}$Se\dotfill & 5.13E$-$06 & 9.28E$-$07 \\
$^{ 24}$Mg\dotfill & 1.80E$-$08 & 2.53E$-$08 & $^{ 76}$Se\dotfill & 9.05E$-$06 & 3.39E$-$08 \\
$^{ 25}$Mg\dotfill & 5.30E$-$09 & 1.05E$-$08 & $^{ 77}$Se\dotfill & 2.77E$-$07 & 7.99E$-$09 \\
$^{ 26}$Mg\dotfill & 4.40E$-$08 & 4.30E$-$08 & $^{ 78}$Se\dotfill & 1.32E$-$07 & 5.41E$-$11 \\
$^{ 27}$Al\dotfill & 3.03E$-$09 & 3.32E$-$09 & $^{ 80}$Se\dotfill & 1.49E$-$10 & 2.87E$-$14 \\
$^{ 28}$Si\dotfill & 5.44E$-$08 & 8.45E$-$08 & $^{ 82}$Se\dotfill & 8.40E$-$16 & 0.00E+00 \\
$^{ 29}$Si\dotfill & 1.11E$-$08 & 1.66E$-$08 & $^{ 79}$Br\dotfill & 2.08E$-$07 & 1.15E$-$08 \\
$^{ 30}$Si\dotfill & 6.82E$-$08 & 6.13E$-$08 & $^{ 81}$Br\dotfill & 1.23E$-$07 & 3.09E$-$09 \\
$^{ 31}$P \dotfill & 2.10E$-$08 & 2.82E$-$08 & $^{ 78}$Kr\dotfill & 7.06E$-$07 & 2.58E$-$07 \\
$^{ 32}$S \dotfill & 1.98E$-$07 & 2.76E$-$07 & $^{ 80}$Kr\dotfill & 2.88E$-$06 & 4.71E$-$08 \\
$^{ 33}$S \dotfill & 3.37E$-$08 & 3.89E$-$08 & $^{ 82}$Kr\dotfill & 1.08E$-$06 & 2.53E$-$08 \\
$^{ 34}$S \dotfill & 2.27E$-$07 & 2.20E$-$07 & $^{ 83}$Kr\dotfill & 2.03E$-$07 & 5.58E$-$09 \\
$^{ 36}$S \dotfill & 4.09E$-$12 & 5.03E$-$12 & $^{ 84}$Kr\dotfill & 3.57E$-$08 & 6.87E$-$12 \\
$^{ 35}$Cl\dotfill & 1.78E$-$07 & 2.03E$-$07 & $^{ 86}$Kr\dotfill & 1.24E$-$11 & 3.48E$-$18 \\
$^{ 37}$Cl\dotfill & 1.05E$-$07 & 1.14E$-$07 & $^{ 85}$Rb\dotfill & 1.40E$-$07 & 4.29E$-$09 \\
$^{ 36}$Ar\dotfill & 1.11E$-$06 & 1.41E$-$06 & $^{ 87}$Rb\dotfill & 4.21E$-$09 & 1.39E$-$15 \\
$^{ 38}$Ar\dotfill & 2.62E$-$07 & 3.01E$-$07 & $^{ 84}$Sr\dotfill & 3.10E$-$07 & 5.87E$-$08 \\
$^{ 40}$Ar\dotfill & 9.90E$-$11 & 1.24E$-$10 & $^{ 86}$Sr\dotfill & 1.29E$-$06 & 6.94E$-$09 \\
$^{ 39}$K \dotfill & 4.07E$-$07 & 4.83E$-$07 & $^{ 87}$Sr\dotfill & 2.92E$-$07 & 2.72E$-$09 \\
$^{ 40}$K \dotfill & 4.98E$-$10 & 5.98E$-$10 & $^{ 88}$Sr\dotfill & 2.74E$-$06 & 3.24E$-$08 \\
$^{ 41}$K \dotfill & 8.14E$-$08 & 8.53E$-$08 & $^{ 89}$Y \dotfill & 3.92E$-$06 & 1.25E$-$08 \\
$^{ 40}$Ca\dotfill & 3.55E$-$06 & 3.93E$-$06 & $^{ 90}$Zr\dotfill & 1.21E$-$04 & 8.04E$-$07 \\
$^{ 42}$Ca\dotfill & 5.23E$-$07 & 6.98E$-$07 & $^{ 91}$Zr\dotfill & 9.96E$-$07 & 6.61E$-$08 \\
$^{ 43}$Ca\dotfill & 1.77E$-$07 & 2.27E$-$07 & $^{ 92}$Zr\dotfill & 7.81E$-$09 & 1.35E$-$10 \\
$^{ 44}$Ca\dotfill & 5.96E$-$06 & 7.17E$-$06 & $^{ 94}$Zr\dotfill & 3.89E$-$16 & 1.97E$-$20 \\
$^{ 46}$Ca\dotfill & 2.44E$-$15 & 1.25E$-$15 & $^{ 96}$Zr\dotfill & 0.00E+00 & 0.00E+00 \\
$^{ 48}$Ca\dotfill & 0.00E+00 & 0.00E+00 & $^{ 93}$Nb\dotfill & 2.05E$-$08 & 2.07E$-$09 \\
$^{ 45}$Sc\dotfill & 1.41E$-$07 & 1.46E$-$07 & $^{ 92}$Mo\dotfill & 7.02E$-$07 & 2.73E$-$07 \\
$^{ 46}$Ti\dotfill & 7.18E$-$07 & 8.29E$-$07 & $^{ 94}$Mo\dotfill & 1.42E$-$08 & 2.32E$-$10 \\
$^{ 47}$Ti\dotfill & 9.24E$-$07 & 1.17E$-$06 & $^{ 95}$Mo\dotfill & 7.96E$-$11 & 1.16E$-$11 \\
$^{ 48}$Ti\dotfill & 8.36E$-$06 & 1.01E$-$05 & $^{ 96}$Mo\dotfill & 3.94E$-$12 & 6.44E$-$14 \\
$^{ 49}$Ti\dotfill & 7.47E$-$07 & 7.64E$-$07 & $^{ 97}$Mo\dotfill & 1.46E$-$12 & 2.09E$-$13 \\
$^{ 50}$Ti\dotfill & 3.35E$-$12 & 7.44E$-$14 & $^{ 98}$Mo\dotfill & 6.96E$-$18 & 0.00E+00 \\
$^{ 50}$V \dotfill & 1.94E$-$09 & 2.97E$-$10 & $^{100}$Mo\dotfill & 0.00E+00 & 0.00E+00 \\
$^{ 51}$V \dotfill & 1.95E$-$06 & 2.23E$-$06 & $^{ 96}$Ru\dotfill & 2.59E$-$11 & 1.58E$-$11 \\
$^{ 50}$Cr\dotfill & 1.41E$-$06 & 1.64E$-$06 & $^{ 98}$Ru\dotfill & 2.86E$-$12 & 6.57E$-$14 \\
$^{ 52}$Cr\dotfill & 1.50E$-$05 & 1.82E$-$05 & $^{ 99}$Ru\dotfill & 1.70E$-$14 & 8.73E$-$16 \\
$^{ 53}$Cr\dotfill & 1.04E$-$06 & 1.22E$-$06 & $^{100}$Ru\dotfill & 7.23E$-$15 & 4.99E$-$16 \\
$^{ 54}$Cr\dotfill & 1.26E$-$08 & 1.86E$-$09 & $^{101}$Ru\dotfill & 2.58E$-$16 & 2.17E$-$17 \\
$^{ 55}$Mn\dotfill & 3.41E$-$06 & 2.88E$-$06 & $^{102}$Ru\dotfill & 9.02E$-$19 & 0.00E+00 \\
$^{ 54}$Fe\dotfill & 3.22E$-$06 & 3.64E$-$06 & $^{104}$Ru\dotfill & 0.00E+00 & 0.00E+00 \\
$^{ 56}$Fe\dotfill & 2.52E$-$03 & 2.92E$-$03 & $^{103}$Rh\dotfill & 4.39E$-$18 & 1.01E$-$19 \\
$^{ 57}$Fe\dotfill & 1.80E$-$04 & 2.13E$-$04 & $^{102}$Pd\dotfill & 3.81E$-$16 & 1.65E$-$17 \\
$^{ 58}$Fe\dotfill & 7.25E$-$08 & 2.18E$-$08 & $^{104}$Pd\dotfill & 3.19E$-$18 & 4.54E$-$21 \\
$^{ 59}$Co\dotfill & 8.61E$-$05 & 1.08E$-$04 & $^{105}$Pd\dotfill & 1.52E$-$20 & 0.00E+00 
\enddata
\end{deluxetable}

\bigskip
\bigskip
\bigskip
\bigskip
\bigskip

\begin{deluxetable}{cccccc}
\tablecaption{Yields of Unstable Isotopes (in Units of $M_\odot$)}
\tablewidth{0pt}
\tablehead{
\colhead{Species} &
\colhead{ST} &
\colhead{FP3} &
\colhead{Species} &
\colhead{ST} &
\colhead{FP3}
}
\startdata
$^{ 22}$Na\dotfill & 5.37E$-$09 & 5.81E$-$09 & $^{ 60}$Fe\dotfill & 1.40E$-$14 & 4.79E$-$16 \\
$^{ 26}$Al\dotfill & 3.27E$-$08 & 2.97E$-$08 & $^{ 56}$Ni\dotfill & 2.50E$-$03 & 2.91E$-$03 \\
$^{ 41}$Ca\dotfill & 8.05E$-$08 & 8.43E$-$08 & $^{ 57}$Ni\dotfill & 1.77E$-$04 & 2.11E$-$04 \\
$^{ 44}$Ti\dotfill & 5.96E$-$06 & 7.17E$-$06 & $^{ 92}$Nb\dotfill & 6.35E$-$09 & 1.34E$-$10 
\enddata
\end{deluxetable}

\end{document}